# Impact of thermal annealing on the interaction between monolayer MoS$_2$ and Au

Stephanie Lough[1], Jesse E. Thompson[1], Darian Smalley[1], Rahul Rao[2*], and Masahiro Ishigami[1]

1. Department of Physics and NanoScience Technology Center, University of Central Florida, Orlando FL USA 32816
2. Materials and Manufacturing Directorate, Air Force Research Laboratory, Wright Patterson Air Force Base, Dayton, OH 45433

*Correspondence – rahul.rao.2@us.af.mil

## ABSTRACT

We have investigated the impact of thermal annealing on the interaction between monolayer MoS$_2$ and Au using Raman spectroscopy. We found MoS$_2$ has two main modes of interactions with the underlying Au being either weakly-coupled or strongly-coupled. The regions strongly-coupled to Au are hybridized to Au, minimally strained, and electron-doped. The weakly-coupled regions are found to be slightly hole-doped with tensile strain of 1.0 %. The observed nanoscale inhomogeneities in doping would result in Au contacts having a large variability in performance. The overall areal coverage of the strongly-coupled regions is not increased by thermal annealing, and the variability in the degree of hybridization increases at annealing temperatures above 100 °C. Our data also show that monolayer MoS$_2$ starts to decouple from Au around 100 °C, becoming fully decoupled above 200 – 250 °C, suggesting that monolayer MoS$_2$ produced by Au-assisted mechanical exfoliation may be more easily transferred off Au at elevated temperatures.

# INTRODUCTION

Next-generation technology nodes are expected to reach the sub-nanometer range via the introduction of ultra-thin and short-channel devices. Silicon (Si) suffers from scalability limits due to short-channel effects and roughness-induced scattering of charge carriers, which greatly reduces their mobility at these size scales [1]. Furthermore, as Si films approach less than 3 nm in thickness, quantum confinement effects can increase the threshold voltage needed to operate the device [2]. $MoS_2$, a two-dimensional (2D) transition metal dichalcogenide (TMD), has been shown to possess carrier mobility an order of magnitude higher than Si at thicknesses less than 3 nm [1]. Yet, their utility is reduced by contact resistance issues that plague 2D TMD devices. In particular, good contacts cannot be made to p-type devices and large variabilities are seen for n-type contacts [3-6].

Experiments have shown that degrees of hybridization and defect generation during the metallization process plays a critically important role in determining the contact resistance [7-11]. Specifically, mechanical transfer of metals results in van der Waals contacts with reduced pinning and enhanced performance [10,11]. Furthermore, varying substrate temperatures can result in n- and p-type devices [6-8] due to the different degrees of defect generation and hybridization during the metallization process. Investigations on monolayer $MoS_2$ produced using gold-assisted mechanical exfoliation [12,13] have shown that the degree of hybridization varies significantly across the sample [14,15], suggesting the interaction between metals and $MoS_2$ plays a large role in the variability of the contact resistance.

We perform a detailed Raman spectroscopy study on monolayer $MoS_2$ on gold to elucidate the impact of hybridization on doping and strain. Thermal annealing is used to remove and deconvolute the impacts of adsorbates. The hybridization results in electron doping but does not introduce strain. Our data also indicate the annealing does not increase the areal coverage of the hybridized regions or affect strain and doping brought on by the gold, but it does introduce more variability in the degree of hybridization. Finally, we find that MoS$_2$ becomes fully decoupled from Au above 250 °C, leading to the possible use of elevated temperatures as a method for transferring mechanically exfoliated monolayer MoS$_2$ from Au onto various substrates for device fabrication.

**RESULTS**

The Raman spectrum from monolayer MoS$_2$ exhibits two characteristic peaks, the in-plane E$_{2g}$ and out-of-plane A$_{1g}$ peaks. When in contact with Au and prior to any annealing, these two MoS$_2$ peaks each appear to be split into two distinct peaks, separated by only a few wavenumbers, as shown in the bottommost spectra in Figures 1a and 1b. The splitting of the A$_{1g}$ peak due to the interaction with Au has been reported [14,15] but the observation of the splitting of E$_{2g}$ has not been previously reported. We label peaks appearing at 378 cm$^{-1}$ and 403 cm$^{-1}$ as the weakly-coupled peaks, $\mathrm{E}_{2g}^{WC}$ and $\mathrm{A}_{1g}^{WC}$, since they are close to the nominal Raman frequencies of monolayer MoS$_2$. We label the shifted satellite peaks at 382 cm$^{-1}$ and 397 cm$^{-1}$ as the strongly-coupled peaks, $\mathrm{E}_{2g}^{SC}$ and $\mathrm{A}_{1g}^{SC}$. These coupling-induced peaks have been attributed to the interaction between Au and MoS$_2$ [14]. Our observation of the split E$_{2g}$ and A$_{1g}$ peaks shows that the MoS$_2$ segregates into regions with weaker and stronger coupling to Au. A previous study [14] has identified weakly-coupled areas to be physically suspended above the underlying Au. We note that the previous studies [14,15] also showed that the sizes of the weakly- and strongly-coupled regions are on the order of few tens to hundreds of nm. Hence, we access both weakly and strongly-coupled regions within our Raman spectral spot (~ 1 µm diameter).

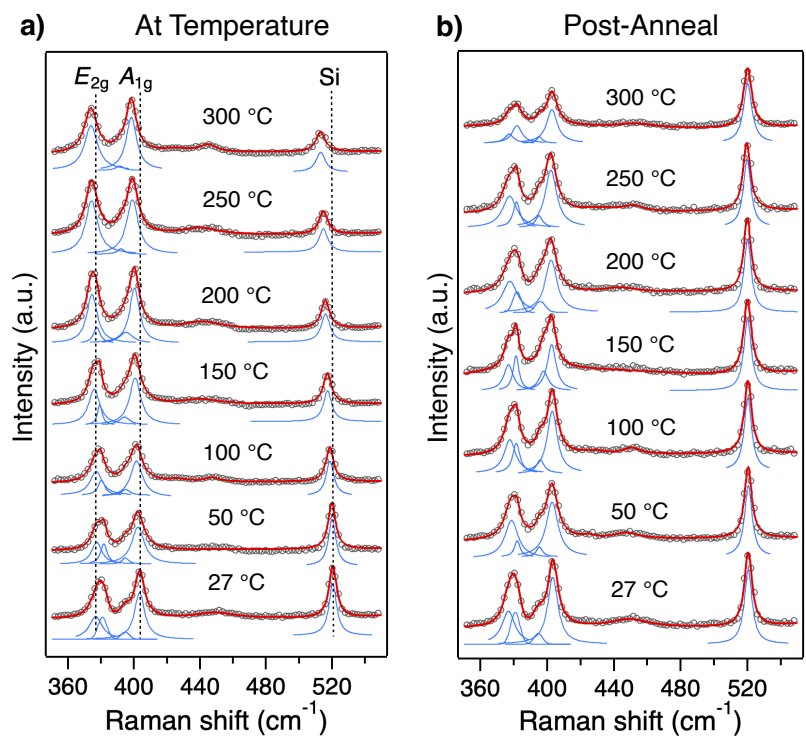

**Figure 1 Temperature-dependent Raman spectroscopy of monolayer MoS$_2$ on Au**. **a)** Spectra measured at a given annealing temperature and **b)** at room temperature post-anneal. Spectra in (a) are normalized with respect to the highest intensity peak and the spectra in (b) are normalized with respect to the intensity of the silicon peak ~ 521 cm$^{-1}$. Overall fits (red curves) are overlaid onto the raw data (black circle data points). Curves are fit with Lorentzian functions and are shown in blue. Black dotted lines indicate nominal peak positions for monolayer MoS$_2$ and Si.

Figure 1a shows representative Raman spectra of monolayer MoS$_2$ collected at the annealing temperatures. The Raman peaks from the MoS$_2$ and Si substrate redshift linearly with increasing annealing temperature, as expected due to lattice anharmonicity effects. The temperature coefficients ($\chi$, slope of the linear redshifts) for the weakly coupled MoS$_2$ and Si peaks are 0.019 ± 0.0016, 0.021 ± 0.0013 and 0.025 ± 0.0018 cm$^{-1}$/K (Figure S1). These coefficients are similar to those reported in the literature for MoS$_2$ and Si [16,17], indicating that monolayer MoS$_2$ is behaving as expected. Figure 1a also shows that the coupling-induced peaks diminish at higher temperatures and nearly completely disappear at temperatures between 200 °C and 250 °C. This indicates that MoS$_2$ is decoupled from Au at higher temperatures. The decoupling can also be more clearly be seen in the intensity ratio between the A$_{1g}$ peak associated with the weakly-coupled regions ($A_{1g}^{WC}$), and the Si peak at 521 cm$^{-1}$ both at elevated temperatures and after cooling down to room temperature as shown in Figure 2a. The observed thermally-induced separation starts at 100 °C and increases with temperature. In addition, Figure 2b also conveys MoS$_2$ delamination by showing strongly-coupled to weakly-coupled peak ratios diminishing at temperature. However, the ratios return upon cooling and indicate that the overall areal coverage of strongly-coupled regions remain constant as the annealing temperature increases.

Figure 1b shows Raman spectra of monolayer MoS$_2$ post-anneal after naturally cooling to room temperature. Annealing does not affect the overall appearance of the spectra once cooled. As such, Figures 1 and 2 indicate that MoS$_2$ becomes decoupled from Au while at high temperatures and then reestablishes the originally observed segregation between weakly- and strongly-coupled regions when cooled. Similar trends in the Raman spectra are observed for MoS$_2$ on Au-coated Si/SiO$_2$ (280 nm oxide layer) substrates (Figure S2). We also note that we do not observe Raman spectral evidence for the 1T' phase of MoS$_2$ as has been reported previously for MoS$_2$ on Au [18].

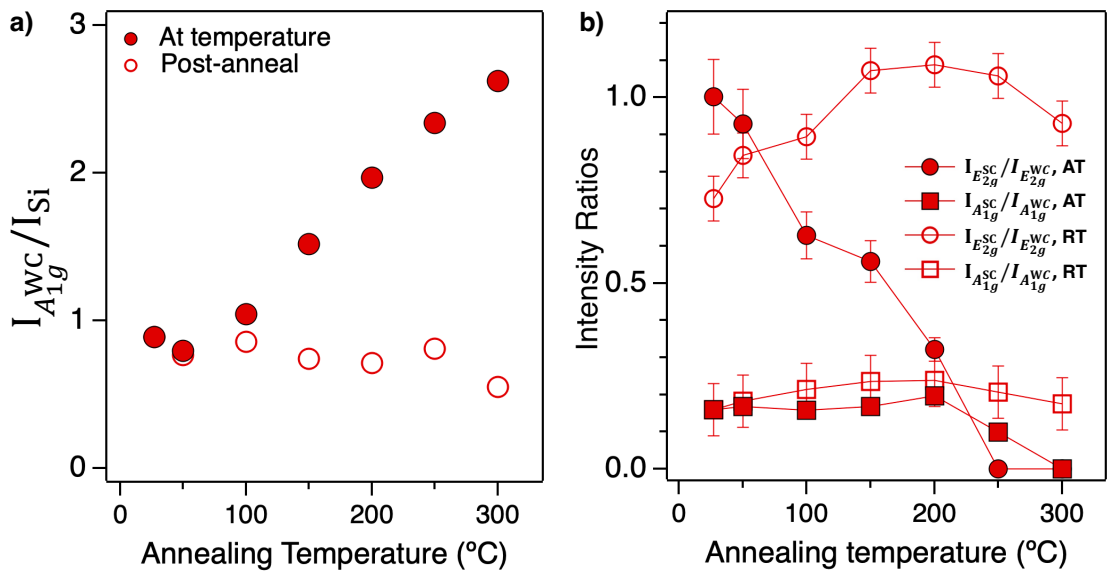

**Figure 2 Raman peak intensity ratios. a)** Ratio of intensities of the $A_{1g}$ peak from the weakly-coupled regions to the silicon peak at temperature and after cooling to room temperature post-anneal, illustrating the delamination of $MoS_2$ from the Au at temperature and the return of intimate contact upon cooling. **b)** Peak intensity ratios of strongly-coupled peaks to weakly-coupled peaks for both $E_{2g}$ and $A_{1g}$ peaks at temperature (AT) and after cooling to room temperature post-anneal (RT). It is evident that the strongly-coupled peaks diminish at higher temperatures, again showing $MoS_2$ delamination, and overall areal coverage of strongly-coupled does not increase upon successive annealing.

In addition to observing coupling between the $MoS_2$ and Au substrate, the Raman spectral features ($A_{1g}$ and $E_{2g}$ peak frequencies) can also be used to deconvolute the effects of strain and doping in monolayer $MoS_2$. The $A_{1g}$ and $E_{2g}$ frequencies have been found to exhibit a quasi-linear dependence on hole and electron doping as well as tensile and compressive strain[19]. The effect of strain and doping can be deconvoluted by plotting the $A_{1g}$ and $E_{2g}$ peak frequencies against each other in a cross-correlation plot [20,21]. The cross-correlation plot for the weakly- and strongly-coupled $MoS_2$ peak frequencies is shown in Figure 3a, with the data points color-coded as a function of annealing temperature. The data in Figure 3 correspond to the post-annealed, room temperature Raman peak frequencies. Note that to generate this plot, we used the peak frequencies of suspended monolayer MoS$_2$ [22-24] as the origin, i.e., corresponding to the unstrained and undoped state. For both the weakly- and strongly-coupled MoS$_2$, the data points line up along constant strains, suggesting that, on the whole, the MoS$_2$ is under tension, but the tensile strains are drastically different for the weakly- and strongly-coupled regions. The strongly-coupled regions have strain values of ~0.2 % while the weakly-coupled regions show a much higher strain value at ~1.0 %. Low strain in the strongly-coupled MoS$_2$ can be understood by similar lattice constants of MoS$_2$ and Au as we found previously[25,26]. Higher strain in the weakly-coupled MoS$_2$ suspended from Au is likely due to MoS$_2$ accommodating the strong-coupling to the underlying gold substrate which has nanometer-scale roughness. A previous study [27] showed that Raman spectroscopy and x-ray diffraction can yield wildly varying results regarding strain values at an MoS$_2$/metal interface compared to the results of our study. Our data suggests that this disagreement could be due, in part, to the inhomogeneity of strain between the weakly- and strongly-coupled regions.

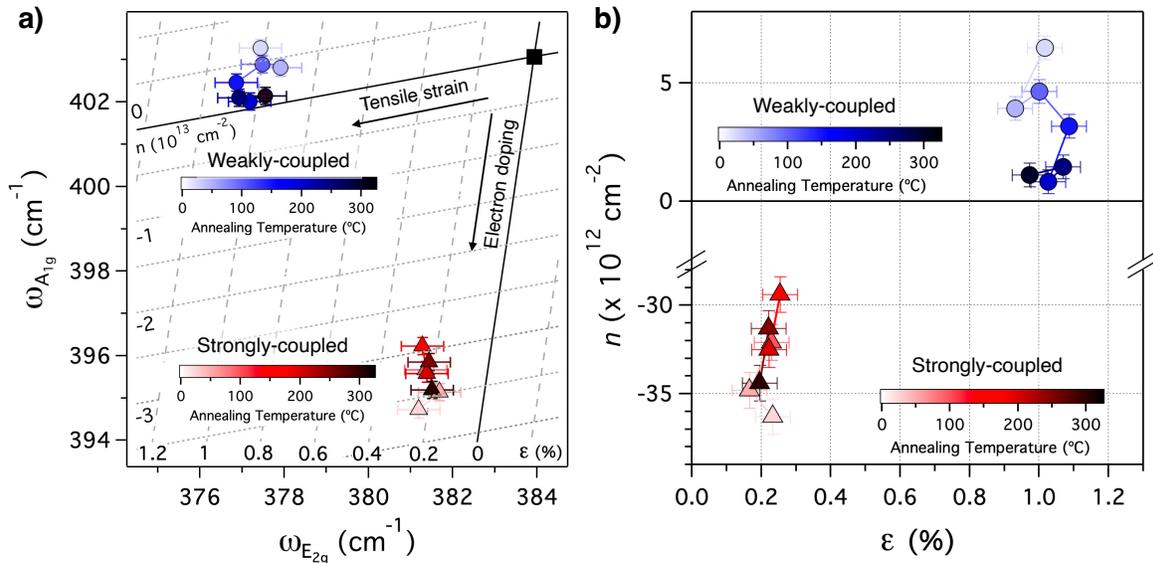

**Figure 3 Strain and doping cross-correlation. a)** Temperature-dependent MoS$_2$ E$_{2g}$ and A$_{1g}$ peak frequencies are plotted against one another to elucidate information on doping and strain. **b)** Strain and doping for each region plotted against one another, showing relatively little change in strain values in each region as a function of temperature. Doping values trend toward the charge-neutrality as the temperature increases, but appears to reach maximum change around 200 °C.

Figure 3b shows the calculated the electron/hole densities (*n*, with positive and negative values corresponding to holes and electrons, respectively) as a function of strain. The weakly-coupled regions are found to be hole-doped as-exfoliated at room temperature with a density of ~ 6.5 x $10^{12}$ cm$^{-2}$. These regions become less hole-doped and trend towards charge-neutral (~1.1 x $10^{12}$ cm$^{-2}$) as the annealing temperatures increases, likely due to desorption of adsorbates at higher annealing temperatures. The strongly-coupled regions are heavily electron-doped as-exfoliated at room temperature, with an electron density of 3.63 × $10^{13}$ cm$^{-2}$. With an increase in annealing temperature, the electron density first decreases to 2.9 × $10^{13}$ cm$^{-2}$, followed by an increase to 3.4 × $10^{13}$ cm$^{-2}$. Thus, thermal annealing does not induce significant overall doping changes in the strongly-coupled regions of the MoS$_2$. The work function of polycrystalline Au is measured to be 5.10 eV in vacuum [28]. Compared to the value of Au, MoS$_2$ has been always found to possess a lower work function. Previous Kelvin probe force microscopy studies found that the work function of CVD-grown monolayer MoS$_2$ ranged from 4.04 eV when measured in vacuum to 4.47 eV measured in an oxygen environment [29]. As such, monolayer MoS$_2$ on Au should be slightly hole-doped based on work functions, contrary to our observation. Therefore, we conclude that the observed electron-doping in the strongly-coupled regions is due to hybridization between MoS$_2$ and Au. This presence of hybridization is also manifested in a redshift of the photoluminescence peak by ~100 meV (Figure S3).

Figure 4 shows the peak widths (full width at half maximum intensity, FWHM) from the weakly- and strongly-coupled regions. Across all annealing temperatures, the peak widths from the strongly-coupled regions are generally lower than those from the weakly-coupled regions, which can be attributed to the higher electron doping levels in the strongly-coupled regions [19]. The widths of the weakly-coupled MoS$_2$ peaks also remain constant between 8 – 10 cm$^{-1}$ within the experimental uncertainty across our annealing temperature range. Since the broadening of the E$_{2g}$ peak is related to the defect density [9], this shows that the weakly-coupled MoS$_2$ is not significantly damaged by the annealing process. Likewise, the $\mathrm{E}_{2g}^{\mathrm{SC}}$ peak in the strongly-coupled region maintains its width throughout the annealing process as shown in Figure 4b. This also indicates that the defect density in the coupled regions does not change substantially during the annealing process. However, the out-of-plane phonon mode $A_{1g}^{SC}$ clearly broadens from ~ 5 to 8 cm$^{-1}$ with increasing annealing temperature, consistent with an increased variability in the degree of hybridization.

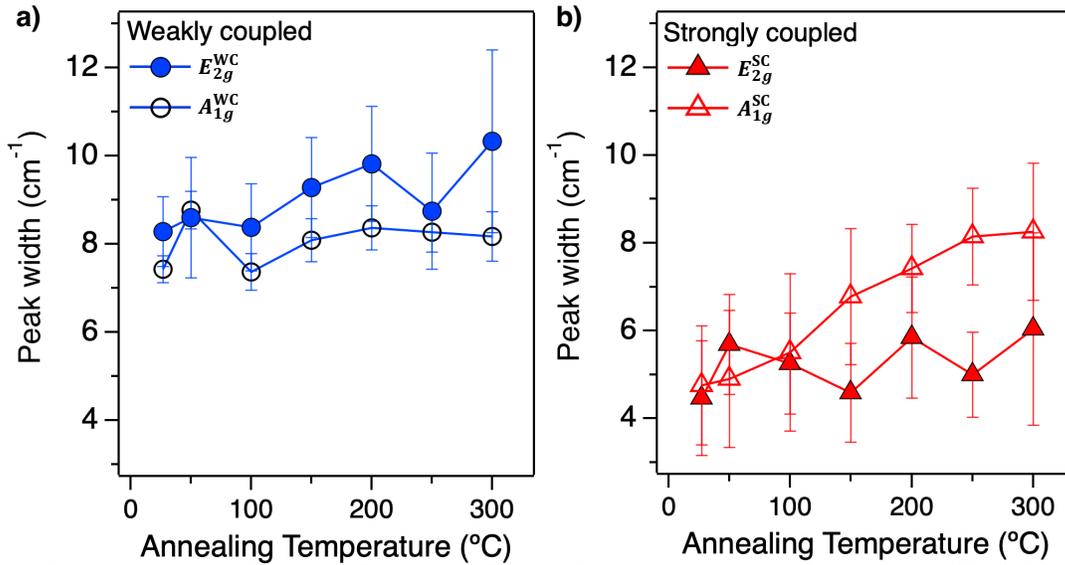

**Figure 4 Raman peak widths as a function of annealing temperature.** Peak widths (FWHM) as a function of temperature for **a)** weakly-coupled regions and **b)** strongly-coupled regions. The results indicate that the monolayer MoS$_2$ does not undergo damage upon successive annealing.

## CONCLUSION

Au-assisted exfoliation results in monolayer MoS$_2$ which has weakly- and strongly-coupled regions as seen in Raman spectra with the splitting of the E$_{2g}$ and A$_{1g}$ peaks. Strongly-coupled regions are found to be hybridized with Au with electron doping at 3.5 × 10$^{13}$ cm$^{-2}$. The weakly-coupled MoS$_2$ regions are found to be slightly hole-doped. In addition, the weakly-coupled regions and tensile strained to ~1.0 %, and this strain reduces to ~0.2% tensile strain in the strongly-coupled MoS$_2$ regions. The observed nanoscale doping inhomogeneities (weakly- and strongly-coupled regions) would result in Au contacts having a large variability in performance. Our results also show that the overall areal coverage of strongly-coupled regions does not increase by thermal annealing while the degree of hybridization becomes more inhomogeneous at annealing temperatures above 100 °C. Therefore, we conclude that thermal annealing is not a viable method for decreasing the inhomogeneity at the interface between monolayer $MoS_2$ and Au systems. Finally, our data also show that monolayer $MoS_2$ starts to decouple from Au at above 100 °C, becoming fully decoupled above 200 °C. This indicates that monolayer $MoS_2$ produced by Au-assisted exfoliation may be more easily transferred off Au at high temperatures.

## METHODS

**Sample Preparation**

Monolayer $MoS_2$ was prepared using gold-assisted exfoliation [12,13] from natural $MoS_2$ crystal, purchased from HQ Graphene. The exfoliation was performed on Au/Cr (40.7 nm/4.2 nm thick) deposited on native oxide $SiO_2$/Si substrates (n-type, 1 – 10 Ohm-cm). Substrates were cleaned by first sonicating in separate acetone and isopropanol baths followed by a reactive ion etch with Trion Reactive Ion and Inductively coupled Plasma at 300 Watts at 13.56 MHz and with $O_2$ rate 98 sccm at 100 mTorr for 500 sec. Metals were deposited using a home-built thermal evaporator at a rate of 2.3 Å/s and 1.6 Å/s for Cr and Au respectively. Immediately after deposition, $MoS_2$ crystals were mechanically exfoliated using blue Nitto tape under ambient conditions and within 1 minute of the vacuum chamber being exposed to air. A limited number of measurements were also performed on $MoS_2$ exfoliated onto Au/Cr-coated (40.6 nm/3.1 nm thick) $SiO_2$/Si substrates (n-type, 0.001 – 0.005 Ohm-cm) with a 280 nm thermal oxide. These additional samples were metallized using the same thermal evaporator (Cr/Au rates of 0.6 Å s$^{-1}$ /1.9Ω Å s$^{-1}$) and the substrates were cleaned, and crystals exfoliated in an identical manner. All samples were stored in air prior to Raman experiments.

**Raman Spectroscopy**

Micro-Raman spectra were acquired with a Renishaw inVia Raman spectrometer utilizing a 514.5 nm laser with a 1 µm spot size. Samples were placed in an in-situ heating vacuum/gas stage (Mikroptic Inc.) where 100 sccm of Ar (99.999 % purity) gas flowed into the sample chamber to keep a positive pressure. For each measurement, samples were heated to the desired temperature, held for 10 minutes, and then allowed to cool naturally back to room temperature. Spectra were fitted with cubic baselines and Lorentzian peaks in Igor Pro to extract peak frequencies, widths, and intensities.

## FUNDING ACKNOWLEDGEMENTS

RR acknowledges funding from the Air Force Office of Scientific Research (AFOSR) grant LRIR 23RXCOR003.

## AUTHOR CONTRIBUTIONS

M.I. and S.L. conceived the experiment. R.R. and S.L. designed and performed experiment. J.E.T., D.S., and S.L. fabricated samples. R.R., M.I., and S.L. analyzed the data. S.L. wrote the manuscript. All authors contributed to the interpretation of experimental results and to the discussion of the manuscript.

## COMPETING INTERESTS

Authors declare no competing interests.

## References

1    English, C. D., Shine, G., Dorgan, V. E., Saraswat, K. C. & Pop, E. Improved Contacts to MoS2 Transistors by Ultra-High Vacuum Metal Deposition (vol 16, pg 3820, 2016). *Nano Lett* **17**, 2739-2739 (2017). https://doi.org/10.1021/acs.nanolett.7b01337

2	Schmidt, M., Lemme, M. C., Gottlob, H. D. B., Driussi, F., Selmi, L. & Kurz, H. Mobility extraction in SOI MOSFETs with sub 1 nm body thickness. *Solid State Electron* **53**, 1246-1251 (2009). https://doi.org/10.1016/j.sse.2009.09.017

3	Chou, A. S. *et al.* Antimony Semimetal Contact with Enhanced Thermal Stability for High Performance 2D Electronics. *Int El Devices Meet* (2021). https://doi.org/10.1109/Iedm19574.2021.9720608

4	O'Brien, K. P. *et al.* Advancing 2D Monolayer CMOS Through Contact, Channel and Interface Engineering. *Int El Devices Meet* (2021). https://doi.org/10.1109/Iedm19574.2021.9720651

5	Knobloch, T., Selberherr, S. & Grasser, T. Challenges for Nanoscale CMOS Logic Based on Two-Dimensional Materials. *Nanomaterials-Basel* **12** (2022). https://doi.org/10.3390/nano12203548

6	Patoary, N. H. *et al.* Improvements in 2D p-type WSe2 transistors towards ultimate CMOS scaling. *Scientific Reports* **13**, 3304 (2023).

7	Wang, Y. *et al.* P-type electrical contacts for 2D transition-metal dichalcogenides. *Nature* **610**, 61-+ (2022). https://doi.org/10.1038/s41586-022-05134-w

8	Wang, Y. & Chhowalla, M. Making clean electrical contacts on 2D transition metal dichalcogenides. *Nat Rev Phys* **4**, 101-112 (2022). https://doi.org/10.1038/s42254-021-00389-0

9	Schauble, K. *et al.* Uncovering the Effects of Metal Contacts on Monolayer MoS2. *Acs Nano* **14**, 14798-14808 (2020). https://doi.org/10.1021/acsnano.0c03515

10	Wang, Y. *et al.* Van der Waals contacts between three-dimensional metals and two-dimensional semiconductors. *Nature* **568**, 70-+ (2019). https://doi.org/10.1038/s41586-019-1052-3

11	Liu, Y. *et al.* Approaching the Schottky-Mott limit in van der Waals metal-semiconductor junctions. *Nature* **557**, 696-+ (2018). https://doi.org/10.1038/s41586-018-0129-8

12	Desai, S. B. *et al.* Gold-Mediated Exfoliation of Ultralarge Optoelectronically-Perfect Monolayers. *Adv Mater* **28**, 4053-4058 (2016). https://doi.org/10.1002/adma.201506171

13	Velicky, M. *et al.* Mechanism of Gold-Assisted Exfoliation of Centimeter-Sized Transition-Metal Dichalcogenide Monolayers. *Acs Nano* **12**, 10463-10472 (2018). https://doi.org/10.1021/acsnano.8b06101

14	Velicky, M. *et al.* Strain and Charge Doping Fingerprints of the Strong Interaction between Monolayer MoS2 and Gold. *J Phys Chem Lett* **11**, 6112-6118 (2020). https://doi.org/10.1021/acs.jpclett.0c01287

15	Pollmann, E. *et al.* Large-Area, Two-Dimensional MoS2 Exfoliated on Gold: Direct Experimental Access to the Metal-Semiconductor Interface. *Acs Omega* **6**, 15929-15939 (2021). https://doi.org/10.1021/acsomega.1c01570

16	Kearney, S. P., Phinney, L. M. & Baker, M. S. Spatially resolved temperature mapping of electrothermal actuators by surface Raman scattering. *J Microelectromech S* **15**, 314-321 (2006). https://doi.org/10.1109/Jmems.2006.872233

17	Najmaei, S., Ajayan, P. M. & Lou, J. Quantitative analysis of the temperature dependency in Raman active vibrational modes of molybdenum disulfide atomic layers. *Nanoscale* **5**, 9758-9763 (2013). https://doi.org/10.1039/c3nr02567e

18	Wu, F. *et al.* Formation of Coherent 1H-1T Heterostructures in Single-Layer MoS2 on Au(111). *Acs Nano* **14**, 16939-16950 (2020). https://doi.org/10.1021/acsnano.0c06014

19	Chakraborty, B., Bera, A., Muthu, D. V. S., Bhowmick, S., Waghmare, U. V. & Sood, A. K. Symmetry-dependent phonon renormalization in monolayer MoS2 transistor. *Phys Rev B* **85** (2012). https://doi.org/10.1103/PhysRevB.85.161403

20	Michail, A., Delikoukos, N., Parthenios, J., Galiotis, C. & Papagelis, K. Optical detection of strain and doping inhomogeneities in single layer MoS2. *Appl Phys Lett* **108** (2016). https://doi.org/10.1063/1.4948357

21  Rao, R. *et al.* Spectroscopic evaluation of charge-transfer doping and strain in graphene/MoS2 heterostructures. *Phys Rev B* **99** (2019). https://doi.org/10.1103/PhysRevB.99.195401

22  Lee, J. U., Kim, K. & Cheong, H. Resonant Raman and photoluminescence spectra of suspended molybdenum disulfide. *2d Materials* **2** (2015). https://doi.org/10.1088/2053-1583/2/4/044003

23  Sahoo, S., Gaur, A. P. S., Ahmadi, M., Guinel, M. J. F. & Katiyar, R. S. Temperature-Dependent Raman Studies and Thermal Conductivity of Few-Layer MoS2. *J Phys Chem C* **117**, 9042-9047 (2013). https://doi.org/10.1021/jp402509w

24  Yan, R. S. *et al.* Thermal Conductivity of Monolayer Molybdenum Disulfide Obtained from Temperature-Dependent Raman Spectroscopy. *Acs Nano* **8**, 986-993 (2014). https://doi.org/10.1021/nn405826k

25  Blue, B. T. *et al.* Metallicity of 2H-MoS2 induced by Au hybridization. *2D Materials* **7**, 025021 (2020). https://doi.org/10.1088/2053-1583/ab6d34

26  Sorkin, V., Zhou, H., Yu, Z. G., Ang, K.-W. & Zhang, Y.-W. The effects of point defect type, location, and density on the Schottky barrier height of Au/MoS2 heterojunction: a first-principles study. *Scientific Reports* **12**, 18001 (2022). https://doi.org/10.1038/s41598-022-22913-7

27  Jernigan, G. G., Fonseca, J. J., Cress, C. D., Chubarov, M., Choudhury, T. H. & Robinson, J. T. Electronic Changes in Molybdenum Dichalcogenides on Gold Surfaces. *J Phys Chem C* **124**, 25361-25368 (2020). https://doi.org/10.1021/acs.jpcc.0c07813

28  Michaelson, H. B. Work Function of Elements and Its Periodicity. *J Appl Phys* **48**, 4729-4733 (1977). https://doi.org/10.1063/1.323539

29  Lee, S. Y. *et al.* Large Work Function Modulation of Monolayer MoS2 by Ambient Gases. *Acs Nano* **10**, 6100-6107 (2016). https://doi.org/10.1021/acsnano.6b01742